\documentclass{article}

\usepackage{amsmath}
\usepackage{amsfonts}
\usepackage{amssymb}
\usepackage{amscd}
\usepackage{amsthm}
\usepackage{latexsym}
\usepackage{times}
\usepackage{mathrsfs}
\usepackage{epsfig}
\usepackage{color}

\usepackage{xy}

\newcommand{\ie}{\emph{i.e.}~}

\def\B{\mathcal B}
\def\C{\mathbb C}
\def\d{\mathrm{d}}
\def\D{\mathcal D}
\def\H{\mathcal H}
\def\Hrond{\mathscr H}
\def\Hi{\mathsf H}
\def\R{\mathbb R}
\def\T{\mathbb T}
\def\U{\mathscr U}

\def\E{\mathcal E}
\def\EE{\mathscr E}
\def\F{\mathscr F}
\def\JJ{\mathscr J}
\def\K{\mathcal K}
\def\M{\mathscr M}
\def\V{\mathscr V}

\newcommand{\bR}{\overline{\R}}
\newcommand{\bRp}{\overline{\R_+}}

\def\({\left(}
\def\){\right)}
\def\[{\left[}
\def\]{\right]}
\def\Rb{[-\infty,\infty]}
\def\slim{\mathop{\hbox{\rm s-}\lim}\nolimits}
\def\e{\mathop{\mathrm{e}}\nolimits}
\def\ee{\varepsilon}
\def\sgn{\mathop{\mathrm{sgn}}\nolimits}
\def\supp{\mathop{\mathrm{supp}}\nolimits}

\def\ltwo{\mathsf{L}^{\:\!\!2}}
\def\lone{\mathsf{L}^{\:\!\!1}}
\def\linf{\mathsf{L}^{\:\!\!\infty}}
\def\dom{\mathcal D}
\def\12{{\textstyle\frac12}}
\def\Tr{\mathop{\mathrm{Tr}}\nolimits}
\def\index{\mathop{\mathrm{index}}\nolimits}

\newtheorem{Theorem}{Theorem}[section]
\newtheorem{Remark}[Theorem]{Remark}
\newtheorem{Assumption}[Theorem]{Assumption}
\newtheorem{Lemma}[Theorem]{Lemma}
\newtheorem{Corollary}[Theorem]{Corollary}

\begin{document}

\title{New formulae for the wave operators for a rank one interaction}

\author{S. Richard$^1\footnote{On leave from Universit\'e de Lyon;
Universit\'e Lyon 1; CNRS, UMR5208, Institut Camille Jordan, 43 blvd du 11 novembre 1918, F-69622 Villeurbanne-Cedex, France.}\ $ and R.
Tiedra de Aldecoa$^2$}
\date{\small}
\maketitle \vspace{-1cm}

\begin{quote}
\emph{
\begin{itemize}
\item[$^1$] Department of Pure Mathematics and Mathematical Statistics,
Centre for Mathematical Sciences, University of Cambridge,
Cambridge, CB3 0WB, United Kingdom
\item[$^2$] Facultad de Matem\'aticas, Pontificia Universidad Cat\'olica de Chile,\\
Av. Vicu\~na Mackenna 4860, Santiago, Chile
\item[] \emph{E-mails:} sr510@cam.ac.uk,
rtiedra@mat.puc.cl
\end{itemize}
  }
\end{quote}

\begin{abstract}
We prove new formulae for the wave operators for a Friedrichs scattering system with
a rank one perturbation, and we derive a topological version of Levinson's theorem
for this model.
\end{abstract}

\section{Introduction and main results}\label{introFrie}
\setcounter{equation}{0}

Let us consider the Hilbert space $\H:=\ltwo(\R)$ with norm $\|\cdot\|$ and scalar
product $\langle\cdot,\cdot\rangle$, and let $H_0\equiv X$ be the operator of
multiplication by the variable, \ie $(H_0f)(x)=xf(x)$ for any
$f\in\dom(H_0)\equiv\H_1$. For $u\in\H$, we also consider the rank one perturbation
of $H_0$ defined by
$$
H_uf:=H_0f+\langle u,f\rangle\;\!u,\quad f \in\H_1.
$$
It is well known that the wave operators $\Omega_\pm:=\slim_{t\to\pm\infty}\e^{iH_ut}\e^{-iH_0t}$ exist and are asymptotically complete, and that the scattering operator $S:=\Omega_+^*\Omega_-$ is a unitary
operator in $\H$. In fact, $S\equiv S(X)$ is simply an operator of multiplication by
a function $S:\R\to\T$, with $\T$ the set of complex numbers of modulus $1$.

A rather explicit formula for the wave operators for this model was proposed in
\cite{GP}. Its expression involves singular integrals that have to be manipulated with
some care. In this Note, we propose a simpler formula for the wave operators, and put
into light a straightforward consequence of it. However, we stress that contrary to \cite{GP}, our formula and its corollary hold only if some additional (but weak) hypotheses on
$u$ are imposed.

To state our results, we need to introduce the even$\;\!$/$\;\!$odd representation of
$\H$. Given any function $m$ on $\R$, we write $m_{\rm e}$ and $m_{\rm o}$ for the
even part and the odd part of $m$. We also set $\Hrond:=\ltwo(\R_+;\C^2)$ and
introduce the unitary map $\U:\H\to\Hrond$ given by
$$
\U f:=\sqrt2
\begin{pmatrix}
f_{\rm e}\\
f_{\rm o}
\end{pmatrix}\quad{\rm and}\quad
\big[\U^*
\big(\begin{smallmatrix}
f_1\\
f_2
\end{smallmatrix}\big)\big](x)
:=\textstyle\frac1{\sqrt2}
\[f_1(|x|)+\sgn(x)f_2(|x|)\],
\quad f\in\H,~\big(\begin{smallmatrix}
f_1\\
f_2
\end{smallmatrix}\big)\in\Hrond,~x\in\R.
$$
Now, observe that if $m$ is a function on $\R$ and $A$ the generator of dilations in
$\H$, then we have
$$
\U m(X)\U^*=
\(\begin{smallmatrix}
m_{\rm e}(X_+)~ & m_{\rm o}(X_+)\\
m_{\rm o}(X_+)~ & m_{\rm e}(X_+)
\end{smallmatrix}\)
\qquad{\rm and}\qquad
\U A\U^*=
\(\begin{smallmatrix}
A_+ & 0\\
0 & A_+
\end{smallmatrix}\),
$$
where $X_+$ is the operator of multiplication by the variable in $\ltwo(\R_+)$, and
$A_+$ the generator of dilations in $\ltwo(\R_+)$, namely $(\e^{itA_+}f)(x):=\e^{t/2}f(\e^tx)$ for $f\in\ltwo(\R_+)$, $x\in\R_+$.

In the sequel we assume that the vector $u$ satisfies the following assumption.

\begin{Assumption}\label{poupette}
The function $u\in\H$ is H\"older continuous with exponent $\alpha>0$. Furthermore,
if $x_0\in\R$ satisfies $u(x_0)=0$ and $1-\int_\R\d y\,|u(y)|^2(x_0-y)^{-1}=0$, then
there exists an exponent $\alpha'>1/2$ such that
$$
|u(x)-u(y)|\le{\rm Const.}\,|x-y|^{\alpha'}
$$
for all $x,y$ near $x_0$.
\end{Assumption}

This assumption clearly implies that $u$ is bounded and satisfies
$\lim_{|x|\to\infty}u(x)=0$. Furthermore, if $u \in \H$ is H\"older continuous with
exponent $\alpha>1/2$, then the previous assumption  holds. On the other hand, under Assumption \ref{poupette}, the operator $H_u$ satisfies the hypotheses of \cite[Sec.~2]{Als80}, and so it has at most a finite number of eigenvalues of
multiplicity one.

Our main result is the following representation of the wave operator $\Omega_-$ in
$\Hrond$.

\begin{Theorem}\label{thmmain}
Let $u$ satisfy Assumption \ref{poupette}. Then, one has
\begin{equation}\label{nice}
\U\Omega_-\U^*=
\(\begin{smallmatrix}
1 & 0\\
0 & 1
\end{smallmatrix}\)
+ \12\(\begin{smallmatrix}
1 & -\tanh(\pi A_+)+i\cosh(\pi A_+)^{-1}\\
-\tanh(\pi A_+)-i\cosh(\pi A_+)^{-1} & 1
\end{smallmatrix}\)
\(\begin{smallmatrix}
S_{\rm e}(X_+)-1 & S_{\rm o}(X_+)\\
S_{\rm o}(X_+) & S_{\rm e}(X_+)-1
\end{smallmatrix}\) + K,
\end{equation}
where $K$ is a compact operator in $\Hrond$.
\end{Theorem}

Let us immediately mention that a similar formula holds for $\Omega_+$. Indeed, by
using $\Omega_+=\Omega_-\overline S(X)$ one gets
$$
\U\Omega_+\U^*=
\(\begin{smallmatrix}
1 & 0\\
0 & 1
\end{smallmatrix}\)
+\12\(\begin{smallmatrix}
1 & \tanh(\pi A_+)-i\cosh(\pi A_+)^{-1}\\
\tanh(\pi A_+)+i\cosh(\pi A_+)^{-1} & 1
\end{smallmatrix}\)
\(\begin{smallmatrix}
\overline{S_{\rm e}}(X_+)-1 & \overline{S_{\rm o}}(X_+)\\
\overline{S_{\rm o}}(X_+) & \overline{S_{\rm e}}(X_+)-1
\end{smallmatrix}\)+K',
$$
where $K'$ is a compact operator in $\Hrond$. We also note that the proof of Theorem
\ref{thmmain} will make clear why the minimal hypothesis $u\in\H$ is not sufficient in
order to prove the claim.

We now state a corollary of the theorem. Its proof is a straightforward consequence
of formula \eqref{nice}, even if it will require the introduction of an algebraic
framework.

\begin{Corollary}\label{surLev}
Let $u$ satisfy Assumption \ref{poupette}. Then $S(\pm\infty)=1$ and the following
equality holds:
$$
\omega(S)=-\;\hbox{number of eigenvalues of $H_u$},
$$
where $\omega(S)$ is the winding number of the continuous map $S:\R\to\T$.
\end{Corollary}

Such a result was already known for more general perturbations but under stronger
regularity conditions \cite{B, Drey} (see also \cite{F,Y} for general informations on
the Friedrichs model). Our result does require neither the differentiability of the scattering matrix nor the differentiability of $u$. Nonetheless, if $S$ is continuously differentiable, then the winding number can also be expressed in terms of an integral involving the (on-shell) time delay operator, which is the logarithmic derivative of
the scattering matrix \cite{T}.

The content of this Note is the following. In Section \ref{Globiboulga} we prove
Formula \eqref{nice} and derive some auxiliary results. In Section \ref{grosbidon} we
give a description of the algebraic framework involved in the proof of the Corollary
\ref{surLev}, which is proved at the end of the section.

\section{Derivation of the new formula}\label{Globiboulga}
\setcounter{equation}{0}

We start by recalling some notations and results borrowed from \cite{Als80} and
\cite{GP}. We shall always suppose that $u$ satisfies Assumption \ref{poupette}.

For $x\in\R$ and $\ee>0$ we set
$$
I_\pm^\ee(x):=\int_\R\d y\,\frac{|u(y)|^2}{x-y\pm i\ee}\,.
$$
The limit $I_\pm(x):=\lim_{\ee\searrow0}I_\pm^\ee(x)$ exists for all $x\in\R$
\cite[Lemma 1.(a)]{Als80}. Furthermore, the set of $x$ such that $I_\pm(x)=1$ is
finite \cite[p.~396]{Als80}. In consequence the expression $[1-I_\pm(x)]^{-1}$ is
well defined for almost every $x\in\R$, and the domain $\D_\pm$ of $[1-I_\pm(X)]^{-1}$
in $\H$ is dense.

Let $\F$ denote the Fourier transform in $\H$, namely
$$
\textstyle
(\F f)(x):=\frac1{\sqrt{2\pi}}\int_\R\d y\,\e^{-ixy}f(y),\quad f\in\H\cap\lone(\R).
$$
Given a Borel function $m$ on $\R$, we set $m(D):=\F^*m(X)\F$. Finally,
$\chi_{(-\infty,0)}$ stands for the characteristic function for the half-line
$(-\infty,0)$.

We are now in a position to recall the formula \cite[eq.~(56b)]{GP} for $\Omega_-$.
One has
$$
\Omega_-=1-2\pi iu(X)\chi_{(-\infty,0)}(D)\overline u(X)[1-I_+(X)]^{-1}
$$
on $\D_+$ (note that we use a convention for the sign $\pm$ of the wave operators
$\Omega_\pm$ which differs from the one of \cite{GP}). So, if
$K:=-2\pi i[u(X),\chi_{(-\infty,0)}(D)]\overline u(X)[1-I_+(X)]^{-1}$, then one gets
on $\D_+$
\begin{align}
\Omega_--1&=-2\pi iu(X)\chi_{(-\infty,0)}(D)\overline u(X)[1-I_+(X)]^{-1}\nonumber\\
&=\chi_{(-\infty,0)}(D)\big\{-2\pi i|u(X)|^2[1-I_+(X)]^{-1}\big\}+K\nonumber\\
&=\chi_{(-\infty,0)}(D)\{S(X)-1\}+K,\label{pantoufle}
\end{align}
by using \cite[Eq.~(66b)]{GP} in the last equality.

In the next lemma, we determine the action of $\chi_{(-\infty,0)}(D)$ in $\Hrond$.
For that purpose, we define $\phi\in C(\Rb;\T)$ by
$\phi(x):=\tanh(\pi x)+i\cosh(\pi x)^{-1}$ for all $x\in\R$.

\begin{Lemma}\label{surR}
One has $\U \chi_{(-\infty,0)}(D)\U^*=\Phi(A_+)$, where
$$
\Phi(A_+):=
\12\(\begin{smallmatrix}
1 & -\overline\phi(A_+)\\
-\phi(A_+) & 1
\end{smallmatrix}\).
$$
\end{Lemma}

\begin{proof}
The usual Hilbert transform $\Hi$ on $\R$ satisfies $\sgn(D)=i\Hi$. Thus
\begin{equation}\label{coco1}
\chi_{(-\infty,0)}(D)=\12\big(1-\sgn(D)\big)=\12(1-i\Hi).
\end{equation}
Using the expression for $i\Hi$ in terms of the generator of dilations in $\H$ given
in \cite[Lemma 3]{KR}, one gets
\begin{equation}\label{coco2}
\U i\Hi\U^*=\(\begin{smallmatrix}
0 & \tanh(\pi A_+)-i\cosh(\pi A_+)^{-1}\\
\tanh(\pi A_+)+i\cosh(\pi A_+)^{-1} & 0
\end{smallmatrix}\).
\end{equation}
The claim follows then from \eqref{coco1} and \eqref{coco2}.
\end{proof}

We now recall some results on the scattering matrix.

\begin{Lemma}\label{deAlsholm}
Let $u$ satisfy Assumption \ref{poupette}. Then the map $S$ belongs to $C(\Rb;\T)$
and satisfies $S(\pm\infty)=1$.
\end{Lemma}

\begin{proof}
The continuity of $S$ follows from \cite[Thm.~1.(i)]{Als80}. The asymptotic equalities $S(\pm\infty)=1$ follow from the formula $S(x)-1=-2\pi i|u(x)|^2[1-I_+(x)]^{-1}$
together with \cite[Lemma~1.(a)]{Als80} and the fact that
$\lim_{|x|\to\infty}|u(x)|^2=0$.
\end{proof}

The last lemma deals with the remainder term $K$ of Formula \eqref{nice}.

\begin{Lemma}\label{operator_K}
Let $u$ satisfy Assumption \ref{poupette}. Then the operator $[u(X),\chi_{(-\infty,0)}(D)]\overline u(X)[1-I_+(X)]^{-1}$, defined on $\D_+$,
extends to a compact operator in $\H$.
\end{Lemma}

\begin{proof}
(i) Define for all $x\in\R$ the function $\psi(x):=\overline u(x)[1-I_+(x)]^{-1}$.
We know that $\overline u$ is bounded and that $\lim_{|x|\to\infty}\overline u(x)=0$.
We also know from \cite[Lemma~1.(a)]{Als80} that $I_+$ is H\"older continuous and that $\lim_{|x|\to\infty}I_+(x)=0$. So, outside any neighbourhood of the finite
set of points where $I_+$ equals $1$, the function $\psi$ is bounded. Furthermore, Assumption \ref{poupette} and \cite[Lemma~1.(c)]{Als80}, imply that $\psi$ is locally square integrable (see also \cite[p.~2423]{GP}). Therefore, $\psi$ can be written as $\psi=\psi_\infty+\psi_2$, with $\psi_\infty\in\linf(\R)$ and $\psi_2\in\ltwo(\R)$ with support in a small
neighbourhood of the points where $I_+$ equals $1$.

We now show the compacity of the operators $[u(X),\chi_{(-\infty,0)}(D)]\psi_\infty(X)$
and $[u(X),\chi_{(-\infty,0)}(D)]\psi_2(X)$.

(ii) Choose a function $\varphi_1\in C^\infty(\R)$ and a function $\varphi_2\in\linf(\R)$
with compact support such that $\varphi_1+\varphi_2 =\chi_{(-\infty,0)}$. Then
$[u(X),\varphi_1(D)]$ is compact due to \cite[Thm.~C]{Cor}, and $[u(X),\varphi_2(D)]$
is Hilbert-Schmidt due to \cite[Thm.~4.1]{Sim05}. So
$$
[u(X),\chi_{(-\infty,0)}(D)]\psi_\infty(X)
=[u(X),\varphi_1(D)]\psi_\infty(X)+[u(X),\varphi_2(D)]\psi_\infty(X)
$$
is a compact operator.

(iii) For each $f\in\H$ and almost every $x\in\R$, define the operator
$$
(K_0f)(x):={\textstyle\frac i{2\pi}}
\int_\R\d y\,\frac{u(x)-u(y)}{y-x}\,\psi_2(y)f(y).
$$
From the Assumption \ref{poupette} we know that
$$
|u(y+x)-u(y)|\le{\rm Const.}\,|x|^{\alpha'},\quad\alpha'>1/2
$$
for each $y\in\supp(\psi_2)$ and each $x\in\R$ with $|x|$ small enough. In particular,
there exists $\delta>0$ such that
\begin{align*}
&4\pi^2\int_{\R^2} \d x\,\d y\,
\bigg|{\textstyle\frac i{2\pi}}\frac{u(x)-u(y)}{y-x}\,\psi_2(y)\bigg|^2\\
&=\int_\R\d y\int_\R\d x\,\frac{|u(y+x)-u(y)|^2}{x^2}\,|\psi_2(y)|^2\\
&=\int_\R\d y\int_{\R\setminus[-\delta,\delta]}\d x\, \frac{|u(y+x)-u(y)|^2}{x^2}\,|\psi_2(y)|^2
+\int_\R\d y\int_{-\delta}^\delta\d x\,\frac{|u(y+x)-u(y)|^2}{x^2}\,|\psi_2(y)|^2\\
&\leq\int_\R\d y\int_{\R\setminus[-\delta,\delta]}\d x\, \frac{4\|u\|^2_\infty}{x^2}\,|\psi_2(y)|^2
+{\rm Const.}\int_\R\d y\int_{-\delta}^\delta\d x\,|x|^{2(\alpha'-1)}\,|\psi_2(y)|^2\\
&<\infty.
\end{align*}
Thus, $K_0$ is a Hilbert-Schmidt operator. Furthermore, we have for $f\in\D_+$ and
almost every $x\in\R$
\begin{align*}
\big\{[u(X),\chi_{(-\infty,0)}(D)]\psi_2(X)f\big\}(x)
=-\textstyle\frac i2\big\{[u(X),\Hi]\psi_2(X)f\big\}(x)
&= -{\textstyle\frac i{2\pi}}\int_\R\d y\,\frac{u(x)-u(y)}{x-y}\,\psi_2(y)f(y)\\
&=(K_0f)(x).
\end{align*}
Therefore, the operator $[u(X),\chi_{(-\infty,0)}(D)]\psi_2(X)$ extends to an
Hilbert-Schmidt operator.
\end{proof}

\begin{proof}[Proof of Theorem \ref{thmmain}]
The operator $K$ extends to a compact operator due to Lemma \ref{operator_K}. So
Equation \eqref{pantoufle} holds on $\H$, and the claim follows from Lemma \ref{surR}.
\end{proof}

\begin{Remark}\label{sigmaess}
The proof of Corollary \ref{surLev} relies on the fact that the range of the wave
operators is the orthocomplement of the subspace spanned by the eigenvectors of $H_u$.
Since the wave operators are complete, such a property holds if and only if $H_u$ has
no singularly continuous spectrum. Now, by using the characterization of the
singular spectrum recalled in \cite[p.~299]{DNY92} and by taking into account Lemmas
1 and 2 of \cite{Als80} (which are valid since $u$ satisfies Assumption \ref{poupette}),
one easily gets that the singular spectrum of $H_u$ only consists of a finite set. So Assumption \ref{poupette} implies the absence of singularly continuous spectrum for $H_u$.
\end{Remark}

\section{Algebraic framework}\label{grosbidon}
\setcounter{equation}{0}

This section is dedicated to the presentation of the algebraic framework leading to
Corollary \ref{surLev}. Since a similar construction already appears in \cite{KR} for
the proof of Levinson's theorem in one dimensional potential scattering, we only
sketch the construction and refer to this reference for more details.

We start by giving the definition of the Mellin transform associated with the
generator of dilations $A_+$ in $\ltwo(\R_+)$ (see \cite[Sec. 2]{J} for a general
presentation when the operator acts in $\ltwo(\R^n)$). Let $\V:\ltwo(\R_+)\to\ltwo(\R)$
be defined by $(\V f)(x):=\e^{x/2}f(\e^x)$ for $x\in\R$, and remark that $\V$ is a
unitary map with adjoint $\V^*$ given by $(\V^*g)(x)=x^{-1/2}g(\ln x)$ for $x\in\R_+$.
Then, the Mellin transform $\M:\ltwo(\R_+)\to\ltwo(\R)$ is defined by $\M:=\F\V$. Its
main property is that it diagonalizes the generator of dilations, namely,
$\M A_+\M^*=X$. Formally, one also has $\M\ln(X_+)\M^*=-D$.

Let us now recall from Remark \ref{sigmaess} that under the Assumption \ref{poupette}
the wave operators $\Omega_\pm$ are isometries with range projection $1-P_{\rm p}$,
where $P_{\rm p}$ is the projection onto the subspace spanned by the finite number
$N$ of eigenvectors of $H_u$. In particular,  $\Omega_-$ is a Fredholm operator with
$\index(\Omega_-)=-\Tr(P_{\rm p})=-N$. Furthermore, we recall that any Fredholm
operator $F$ in $\H$ is invertible modulo a compact operator, that is, its image
$q(F)$ in the Calkin algebra $\B(\H)/\K(\H)$ is invertible.

Now, assume that $\Omega_-$ belongs to a norm-closed subalgebra $\E$ of $\B(\H)$
containing $\K(\H)$. Moreover, assume that  $\E/\K(\H)$ is isomorphic to
$C\big(\mathbb S;M_2(\C)\big)$, the algebra of continuous functions over the circle
with values in the $2\times2$ matrices. Then, viewing $q(\Omega_-)$ as such a function,
we can take pointwise its determinant to obtain a non-vanishing function over the
circle. The winding number of that latter function can be related to the index of
$\Omega_-$; this is essentially the content of Corollary \ref{surLev}.

The choice of $\E$ is suggested by the formula obtained in Theorem \ref{thmmain}.
Indeed, we consider the closure $\EE$ in $\B(\Hrond)$ of the algebra generated by
elements of the form $\varphi(A_+)\psi(X_+)$, where $\varphi$ is a continuous function
on $\R$ with values in $M_2(\C)$ which converges at $\pm\infty$, and $\psi$ is a
continuous function on $\R_+$ with values in $M_2(\C)$ which converges at $0$ and at $+\infty$. Stated differently, $\varphi\in C\big(\bR;M_2(\C)\big)$ with
$\bR=[-\infty,\infty]$, and $\psi\in C\big(\bRp;M_2(\C)\big)$ with $\bRp=[0,\infty]$.
Let $\JJ$ be the norm closed algebra generated by $\varphi(A_+)\psi(X_+)$ with
functions $\varphi$ and $\psi$ for which the above limits vanish. Then, $\JJ$ is an
ideal in $\EE$, and the same algebras are obtained if $\psi(X_+)$ is replaced by
$\eta(\ln(X_+))$ with $\eta \in C\big(\bR;M_2(\C)\big)$ or
$\eta \in C_0\big(\R;M_2(\C)\big)$, respectively.

These algebras have already been  studied in \cite{Georgescu} in a different context.
The authors constructed them in terms of the operators $X$ and $D$ on $\ltwo(\R,E)$,
with $E$ an auxiliary Hilbert space, possibly of infinite dimension. In that situation, $\varphi$ and $\eta$ are norm continuous functions on $\bR$ with values in $\K(E)$.
The isomorphism between our algebras and the algebras introduced in
\cite[Sec.~3.5]{Georgescu}, with $E=\C^2$, is given by the Mellin transform $\M$, or
more precisely by $\M\otimes 1$, where $1$ is identity operator in $M_2(\C)$. For that reason, we shall freely use the results obtained in that reference, and refer to it
for the proofs. For instance, it is proved that $\JJ=\K(\Hrond)$, and an explicit
description of the quotient $\EE/\JJ$ is given, which we specify now in our context.

To describe the quotient $\EE/\JJ$, we consider the square $\square:=\bRp\times\bR$,
whose boundary $\partial\square$ is the union of four parts:
$\partial\square\equiv B_1\cup B_2\cup B_3\cup B_4$, with $B_1:=\{0\}\times\bR$, $B_2:=\bRp\times\{\infty\}$, $B_3:=\{\infty\}\times\bR$, and
$B_4:=\bRp\times\{-\infty\}$. It is proved in \cite[Thm.~3.20]{Georgescu} that
$\EE/\JJ$ is isomorphic to $C\big(\partial\square;M_2(\C)\big)$. This algebra can be
seen as the subalgebra of
\begin{equation}\label{gros}
C\big(\bR;M_2(\C)\big)\oplus C\big(\bRp;M_2(\C)\big)\oplus C\big(\bR;M_2(\C)\big)
\oplus C\big(\bRp;M_2(\C)\big)
\end{equation}
given by elements $\gamma\equiv(\gamma_1,\gamma_2,\gamma_3,\gamma_4)$ which coincide
at the corresponding end points, that is, $\gamma_1(\infty)=\gamma_2(0)$, $\gamma_2(\infty)=\gamma_3(\infty)$, $\gamma_3(-\infty)=\gamma_4(\infty)$, and
$\gamma_4(0)=\gamma_1(-\infty)$. Furthermore, for any
$\varphi\in C\big(\bR;M_2(\C)\big)$ and $\psi \in C\big(\bRp;M_2(\C)\big)$, the image
of $\varphi(A_+)\psi(X_+)$ through the quotient map
$q:\EE\to C\big(\partial\square;M_2(\C)\big)$ is given by $\gamma_1 = \varphi\psi(0)$, $\gamma_2=\varphi(\infty)\psi$, $\gamma_3=\varphi\psi(\infty)$ and
$\gamma_4=\varphi(-\infty)\psi$.

From what precedes we deduce that the subalgebra $\E$ of $\B(\H)$, defined by $\E:=\U^*\EE\U$, contains the ideal of compact operators on $\H$ and that the
quotient $\E/\K(\H)$ is isomorphic to
$C\big(\partial\square;M_2(\C)\big)\cong C\big(\mathbb S;M_2(\C)\big)$. We are thus in
the framework defined above, and the for any invertible element $\gamma$ of $C\big(\partial\square;M_2(\C)\big)$, the winding number of its pointwise determinant
is a well-defined quantity. So we are ready to give the proof of Corollary
\ref{surLev}.

\begin{proof}[Proof of Corollary \ref{surLev}]
We know from Theorem \ref{thmmain} and Lemma \ref{deAlsholm} that
$\U\Omega_-\U^*\in\EE$, or equivalently that $\Omega_-\in\E$. Due to Formula
\eqref{nice}, the element $\gamma$ belonging to \eqref{gros} and associated with
$q(\Omega_-)$ is given by suitable restrictions of the function
$\Gamma:\bRp\times\bR\to M_2(\C)$, where
$$
\Gamma(x,y):=1+\Phi(y)
\(\begin{smallmatrix}
s_{\rm e}(x)-1 & s_{\rm o}(x)\\
s_{\rm o}(x) & s_{\rm e}(x)-1
\end{smallmatrix}\)
=
\12
\(\begin{smallmatrix}
s_{\rm e}(x)-\overline\phi(y)s_{\rm o}(x)+1
& ~s_{\rm o}(x)-\overline\phi(y)[s_{\rm e}(x)-1]\\
s_{\rm o}(x)-\phi(y)[s_{\rm e}(x)-1]
& ~s_{\rm e}(x)-\phi(y)s_{\rm o}(x)+1
\end{smallmatrix}\).
$$
Namely, $\gamma_1=\Gamma(0,\;\!\cdot\;\!)$, $\gamma_2=\Gamma(\;\!\cdot\;\!,+\infty)$,
$\gamma_3=\Gamma(+\infty,\;\!\cdot\;\!)$, and  $\gamma_4=\Gamma(\;\!\cdot\;\!,-\infty)$.
The pointwise determinants of these functions are easily calculated by using the identity $\phi(\pm\infty)=\pm1$: one gets $\det\gamma_1(y)=s_{\rm e}(0)$, $\det\gamma_2(x)=s(-x)$,
$\det\gamma_3(y)=1$ and $\det\gamma_4(x)=s(x)$.

The precise relation between the winding number of the map
$\det\gamma:\partial\square\to\T$ and the index of $\Omega_-$ has been described in \cite[Prop.~7]{KR}. However, the algebra corresponding to $\EE$ in that reference was
defined in terms of the operators $A_+$ and $B_+=\12\ln\big((D^2)_+\big)$ which satisfy
the relation $[iA_+,B_+]=-1$. In our case, the algebra $\EE$ has been constructed
with the operators $A_+$ and $\ln(X_+)$ which satisfy the relation $[iA_+,\ln(X_+)]=1$. Therefore, in order to apply \cite[Prop.~7]{KR} in our setting, one previously needs to
apply the automorphism of $C\big(\bR;M_2(\C)\big)$ defined by $\widetilde \eta(x):=\eta(-x)$ for all $x\in\R$, or equivalently the automorphism of $C\big(\bRp;M_2(\C)\big)$  defined
by $\widetilde\psi(x):=\psi(x^{-1})$ for all $x \in \R_+$. Therefore the pointwise determinants of the function $\widetilde\gamma_j$ associated with $q(\Omega_-)$ are
$\det\widetilde\gamma_1(y)=1$, $\det\widetilde\gamma_2(x)=s(-x^{-1})$,
$\det\widetilde\gamma_3(y)=s_{\rm e}(0)$ and $\det\widetilde\gamma_4(x)=s(x^{-1})$.

Now, \cite[Prop.~7]{KR} reads  $\omega(\det\widetilde\gamma)=\index(\Omega_-)=-N$, where
$N$ is the number of eigenvalues of $H_u$. The convention used in that reference for the calculation of the winding number implies that the contribution of
$x\mapsto\det \widetilde\gamma_2(x)$ is from $x=0$ to $x=+\infty$ and the contribution
of $x\mapsto \det\widetilde\gamma_4(x)$ is from $x=+\infty$ to $x=0$. This corresponds to
the calculation of the winding number of $x\mapsto \det\big(S(x)\big)$, from $x=-\infty$
to $x=+\infty$. Since the contributions of $\det \widetilde\gamma_1$ and
$\det\widetilde\gamma_3$ are null because these terms are constant, the claim is proved.
\end{proof}

\section*{Acknowledgements}

S. Richard is supported by the Swiss National Science Foundation.
R. Tiedra de Aldecoa is partially supported by N\'ucleo Cient\'ifico ICM P07-027-F ``Mathematical Theory of Quantum and Classical Magnetic Systems" and by the Chilean
Science Fundation Fondecyt under the Grant 1090008.



\begin{thebibliography}{00}

\bibitem{Als80} P. Alsholm, \emph{Inverse scattering theory for perturbations of rank one}, Duke Math. J. {\bf 47} no. 2 (1980), 391--398.

\bibitem{B} V.S. Buslaev, \emph{Spectral identities and the trace formula in the
Friedrichs model}, in Spectral theory and wave processes,  43--54.
Consultants Bureau Plenum Publishing Corporation, New York, 1971.

\bibitem{Cor} H.O. Cordes,
\emph{On compactness of commutators of multiplications and convolutions,
and boundedness of pseudodifferential operators},
J. Funct. Anal. {\bf 18} (1975), 115--131.

\bibitem{Drey} T. Dreyfus,
\emph{The determinant of the scattering matrix and its relation to
the number of eigenvalues}, J. Math. Anal. Appl. {\bf 64} no. 1 (1978), 114--134.

\bibitem{DNY92} Dyn$'$kin, E. M. and Naboko, S. N. and Yakovlev, S. I.,
\emph{A finiteness bound for the singular spectrum in a selfadjoint
{F}riedrichs model}, St. Petersburg Math. J. {\bf 3} no. 2 (1992), 299--313.

\bibitem{F} L.D. Faddeev, \emph{On a model of Friedrichs in the theory of perturbations of the continuous spectrum}, in American Mathematical Society Translations. Series 2, Vol. 62: Five papers on functional analysis, 177--203, American Mathematical Society, Providence, R.I. 1967.

\bibitem{Georgescu} V. Georgescu, A. Iftimovici,
\emph{$C^*$-algebras of quantum Hamiltonians}, in Operator Algebras
and Mathematical Physics, Conference Proceedings: Constan\c{t}a
(Romania) July 2001, 123--167, Theta Foundation, 2003.

\bibitem{GP} M.A. Grubb, D.B. Pearson, \emph{Derivation of the wave and
scattering operators for an interaction of rank one},
J. Mathematical Phys. {\bf 11} (1970), 2415--2424.

\bibitem{J} A. Jensen, \emph{Time-delay in potential scattering theory, some "geometric"\ results}, Comm. Math. Phys.  {\bf 82} no. 3 (1981/82), 435--456.

\bibitem{KR} J. Kellendonk, S. Richard, \emph{On the structure of the wave
operators in one dimensional potential scattering}, to appear in Mathematical Physics Electronic Journal.

\bibitem{Sim05}
B.~Simon, \emph{Trace ideals and their applications},
Mathematical Surveys and Monographs {\bf 120},
American Mathematical Society, Providence, RI, second edition, 2005.

\bibitem{T} R. Tiedra de Aldecoa, \emph{Time delay for dispersive systems in
quantum scattering theory}, mp\_arc 08-179.

\bibitem{Y} D.R. Yafaev, \emph{Mathematical scattering theory. General theory},
Translations of Mathematical Monographs, 105. American Mathematical Society, Providence,
RI, 1992.
\end{thebibliography}
\end{document}